\documentclass[aps,prl,twocolumn]{revtex4-1}

\usepackage{amsmath}
\usepackage{mathtools} 
\usepackage{amssymb}
\usepackage{amsthm}
\usepackage{bm,upgreek} 
\usepackage{upgreek} 
\usepackage{xcolor}
\usepackage{graphicx}
\usepackage{epstopdf}
\usepackage[colorlinks,linkcolor=blue,anchorcolor=blue,citecolor=blue,urlcolor=blue]{hyperref}
\usepackage{tikz}
\usepackage[compat=1.1.0]{tikz-feynman}

\newcommand{\nn}{{\nonumber}}
\newcommand{\bea}{\begin{eqnarray}}
\newcommand{\eea}{\end{eqnarray}}
\newcommand{\ie}{\textit{i.e.{ }}}

\newcommand{\up}{\uparrow}
\newcommand{\dn}{\downarrow}

\newcommand{\av}[1]{\left\langle #1 \right\rangle}

\newcommand{\md}{\mathrm{d}}
\newcommand{\me}{\mathrm{e}}
\newcommand{\w}{\omega}

\begin{document}

\title{Anisotropic scattering caused by apical oxygen vacancies in thin films of overdoped high-temperature cuprate superconductors}
\author{Da Wang}\email{dawang@nju.edu.cn}
\author{Jun-Qi Xu}
\author{Hai-Jun Zhang}
\author{Qiang-Hua Wang}\email{qhwang@nju.edu.cn}
\affiliation{National Laboratory of Solid State Microstructures $\&$ School of Physics, Nanjing University, Nanjing 210093, China}
\affiliation{Collaborative Innovation Center of Advanced Microstructures, Nanjing University, Nanjing 210093, China}

\begin{abstract}
There is a hot debate on the anomalous behavior of superfluid density $\rho_s$ in overdoped La$_{2-x}$Sr$_x$CuO$_4$ films in recent years. The linear drop of $\rho_s$ at low temperatures implies the superconductors are clean, but the linear scaling between $\rho_s$ (in the zero temperature limit) and the transition temperature $T_c$ is a hallmark of the dirty limit in the Bardeen-Cooper-Schrieffer (BCS) framework [\href{https://dx.doi.org/10.1038/nature19061}{I. Bozovic {\it et al.}, Nature {\bf 536}, 309 (2016)}]. This dichotomy motivated exotic theories beyond the standard BCS theory. We show, however, that such a dichotomy can be reconciled naturally by the role of increasing anisotropic scattering caused by the apical oxygen vacancies. Furthermore, the anisotropic scattering also explains the ``missing'' Drude weight upon doping in the optical conductivity, as reported in the THz experiment [\href{https://dx.doi.org/10.1103/PhysRevLett.122.027003}{F. Mahmood {\it et al.}, Phys. Rev. Lett. {\bf 122}, 027003 (2019)}].
Therefore, the overdoped cuprates can actually be described consistently by the $d$-wave BCS theory with the unique anisotropic scattering.
\end{abstract}

\maketitle

\emph{Introduction}.---
For a long time, the overdoped cuprate superconductors were believed to be described quite well by the Bardeen-Cooper-Schrieffer (BCS) theory \cite{Lee2006,Keimer2015}. However, in 2016, such a belief was challenged by the measurement of the superfluid density $\rho_s$ using mutual inductance technique on a large amount of high quality overdoped La$_{2-x}$Sr$_x$CuO$_4$ (LSCO) films \cite{Bozovic2016}. Two seemingly contradicting results were reported: $\rho_s(0)-\rho_s(T)\propto T$ and $\rho_s(0)\propto T_c$ where $\rho_s(0)$ is the zero temperature value of $\rho_s(T)$ and $T_c$ is the transition temperature. Within the conventional BCS theory, the former scaling law indicates the d-wave superconductors are very clean, but the latter is a result of dirty BCS superconductors \cite{AGD1963}.
This dichotomy regarding the dirtiness/cleanness of the overdoped cuprates motivated exotic theoretical investigations \cite{Bozovic2019,Zaanen2016,Dynes2018,Z2tJ2018,DW2018,Holography2020,Phillips2020,Li2021}.

The superfluid density has also been measured by THz optical conductivity experiment \cite{Mahmood2019}, and is quantitatively consistent with the mutual inductance measurement \cite{Bozovic2016}. In the meantime, the Drude fitting of the optical conductivity $\sigma_1(\nu)$ indicates the dirty limit \cite{Lee-Hone2017,Lee-Hone2018}. So the same dichotomy also exists in optical measurements.
Moreover, as yet another puzzle, $\sigma_1(\nu\to0)$ should be identical to, but is fitted to be significantly smaller than the dc conductivity $\sigma_{dc}$ \cite{Mahmood2019}.
This superficial loss of Drude weight seems to increase with overdoping.

In this Letter, we propose a scenario to resolve all of the above mysteries. We realize that in addition to the conventional isotropic scattering rate $\Gamma_s$ \cite{Lee-Hone2017,Lee-Hone2018}, the apical oxygen vacancies cause an anisotropic scattering rate $\Gamma_d\cos^2(2\theta)$, with $\theta$ the azimuthal angle of the quasiparticle momentum relative to the antinodal direction. Since the low-energy nodal quasiparticles are barely affected by $\Gamma_d$, they reduce the superfluid density linearly in temperature if in addition $\Gamma_s\rightarrow 0$. But the total scattering rate $\Gamma_\theta = \Gamma_s+\Gamma_d\cos^2(2\theta)$ determines the typical behavior $\rho_s(0)\propto T_c$ in the dirty limit. Meanwhile, the strong anisotropy in $\Gamma_\theta$ causes a continuous distribution of Lorentzian components in $\sigma(\nu)$, so that $\sigma_1(\nu\to0)$ would be underestimated by extrapolation from finite frequencies if a single isotropic scattering rate were assumed instead \cite{Mahmood2019}.

\begin{figure}
\includegraphics[width=0.6\linewidth]{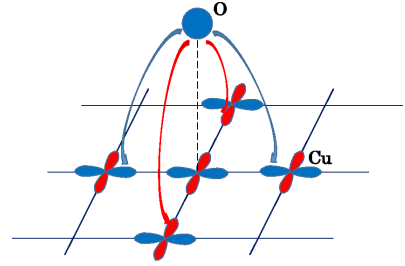}
\caption{Illustration of apical oxygen coupling to (effective) $d_{x^2-y^2}$-ortibals in the CuO$_2$ plane. Hoppings with opposite signs are labeled by blue and red colors, respectively.}
\label{fig:scheme}
\end{figure}

\emph{Oxygen vacancies and ``cold spot'' model}.---
From both early \cite{Ovacancy1988,Sato2000} and recent \cite{Kim2017} studies, apical oxygen vacancies were found to be an important factor to prevent obtaining (high-quality) overdoped LSCO samples. Therefore, the effect of apical oxygen vacancies deserves study in overdoped cuprates carefully. This is another motivation of the present work.

In Fig.~\ref{fig:scheme}, we plot a configuration of one apical oxygen atom above the CuO$_2$-plane (only Cu-sites plotted). Notice that since the carriers have $d_{x^2-y^2}$-like orbital content \cite{ZR1988}, there is no coupling between the oxygen and the $d_{x^2-y^2}$-orbital just below it. Therefore, the leading couplings are with the next nearest neighboring sites, giving the hopping integrals switching signs alternatively, as shown in Fig.~\ref{fig:scheme}. First, if there is no oxygen vacancy, the second order process of these out-of-plane hoppings gives an additional band dispersion $-4\kappa(\cos k_x-\cos k_y)^2$ where $\kappa=t_o^2/\varepsilon_o$ with $t_o$ and $\varepsilon_o$ the hopping amplitude and energy level distance between the apical oxygen and $d_{x^2-y^2}$ orbitals \cite{Xiang1996}. Such a dispersion has already enters the carrier band. Next, we consider one oxygen vacancy at the origin. Relative to the uniform background, the vacancy now leads to an additional term $H_i=\sum_{\delta\delta'\sigma}\kappa f_\delta f_{\delta'}c_{\delta\sigma}^\dag c_{\delta'\sigma}$ where $f_\delta=1(-1)$ for $\delta=\pm\hat{x}(\pm\hat{y})$ and $\sigma$ denotes spin, which gives rise to the following scattering Hamiltonian
\begin{align}\label{eq:scatteringV}
H_i=\sum_{kk'\sigma}\frac{4\kappa}{N}(\cos k_x-\cos k_y)(\cos k_x'-\cos k_y')c_{k\sigma}^\dag c_{k'\sigma}
\end{align}
where $k$ and $k'$ are momentums, and $N$ is the number of copper atoms.
Since the oxygen vacancies are out-of-plane and only couple to next nearest neighboring $d_{x^2-y^2}$-orbitals, the scattering strength is anticipated to be small.
Under the Born approximation \cite{AGD1963}, $H_i$ gives a scattering rate $\Gamma(k_x,k_y)\propto\Gamma_d(\cos k_x-\cos k_y)^2/4$. For analytical convenience but without loss of qualitative physics, throughout this work, we further assume circular fermi surface and use wide band approximation, so that the scattering rate is only angle-dependent, \ie
\begin{align}
\Gamma_\theta=\Gamma_s+\Gamma_d\cos^2(2\theta)
\end{align}
where we have added the isotropic scattering rate $\Gamma_s$ arising from, {\it e.g.}, in-plane impurities. In the following calculations, $\Gamma_s$ and $\Gamma_d$ are our model parameters.
In fact, this kind of scattering rate has been proposed to explain the transport phenomena in underdoped cuprates, called ``cold spot'' model \cite{Ioffe1998}, where the $\Gamma_d$-term is attributed to the pair fluctuations or interaction effects, which hence is expected to decrease upon doing. {But here, in our case, the $\Gamma_d$-term is caused by oxygen vacancies and should increase upon doping, according to both early \cite{Ovacancy1988,Sato2000} and recent \cite{Kim2017} experiments.}

\begin{figure}
\includegraphics[width=\linewidth]{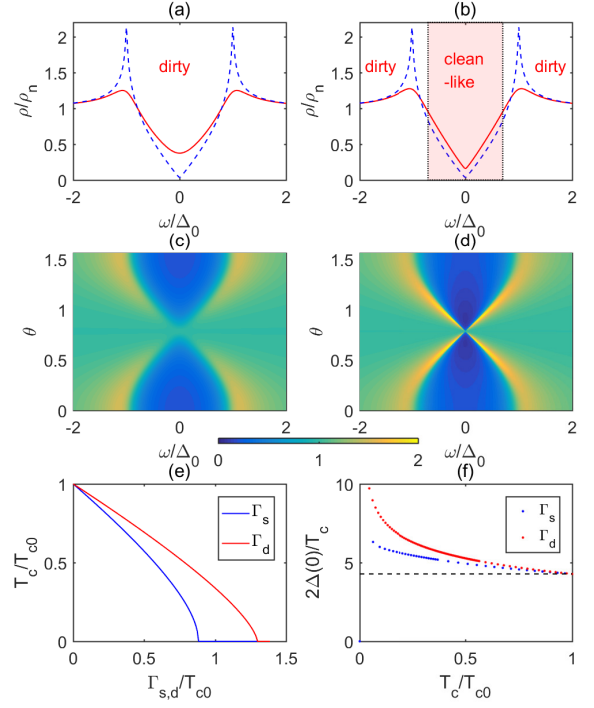}
\caption{(a) The DOS $\rho$ versus $\omega$ in the case of pure isotropic scattering $\Gamma_s=0.2\Delta_0$ (solid line). For comparison, the dashed line shows the DOS in the clean limit. (b) The same as (a) but in the case of pure anisotropic scattering $\Gamma_d=0.2\Delta_0$. The shaded region highlights the similarity to the clean limit. (c) and (d) are angle-dependent PDOS in the presence of scattering corresponding to (a) and (b), respectively. (e) $T_c/T_{c0}$ versus $\Gamma_s/T_{c0}$ (blue) and $\Gamma_d/T_{c0}$ (red) for the two types of scatterings. (f) The ratio $2\Delta_0(0)/T_c$ versus $T_c/T_{c0}$ (red and blue dots) for the two cases considered in (e), in comparison to the clean limit result (dashed line).}
\label{fig:dos}
\end{figure}

We now look at the effect of the anisotropic scattering $\Gamma_d$ on single particle excitations. By choosing the d-wave pairing function $\Delta(\theta)=\Delta_0\cos(2\theta)$ and fixing $\Gamma_{s,d}=0.2\Delta_0$, we calculate the density of states (DOS) $\rho(\w)=\int\frac{\md\theta}{2\pi}\rho(\w,\theta)$, where $\rho(\w,\theta)$ is the partial DOS (PDOS) given by
\begin{align}
\rho(\w,\theta)=\frac{\+N}{\pi}\int\md\varepsilon \left[ \frac{\Gamma_\theta}{(\w-E_\theta)^2+\Gamma_\theta^2} + \frac{\Gamma_\theta}{(\w+E_\theta)^2+\Gamma_\theta^2}\right]
\end{align}
where $\+N$ is the normal state DOS and $E_\theta=\sqrt{\varepsilon^2+\Delta(\theta)^2}$. $\rho(\w)$ and $\rho(\w,\theta)$ are plotted in Fig.~\ref{fig:dos}(a)-(d).
{In the calculations, we perform the integrals over $\varepsilon$ and $\theta$ using standard numerical integral technique.}
Fig.~\ref{fig:dos}(a) is the typical behavior of the conventional d-wave superconductor with isotropic scattering rate: zero energy cusp and coherence peak are both smeared out by isotropic scattering rate. Differently, the $\Gamma_d$-scattering shown in Fig.~\ref{fig:dos}(b) only suppresses the coherence peak but does not change the zero energy cusp, as a result of the vanishing of scattering rate for nodal quasiparticles. Accordingly, we can divide the energy space into two regions: lower energy ``clean'' one and higher energy ``dirty'' one. The above picture is also seen clearly from the PDOS in Fig.~\ref{fig:dos}(d), where the spectral becomes smeared out for high energy quasiparticles (near the antinodes) but remains sharp for low energy ones (near the nodes).

The dirty BCS theory for d-wave superconductors \cite{Alloul2009} is a direct generalization of the Abrikosov-Gor'kov (AG) theory \cite{AG1961}, since the potential scattering here causes pair-breaking which is similar to the magnetic impurities in s-wave superconductors.
The gap $\Delta_0$ is determined by the self-consistent equation
\begin{align}
1=\lambda T\sum_{\w_n>0}^{\Omega}\int\md\theta\frac{\phi_\theta^2}{\sqrt{\w_n^2+2|\w_n|\Gamma_{\theta}+\Gamma^2_{\theta}+\Delta_0^2\phi_\theta^2}}
\end{align}
where $\lambda$ is the BCS coupling constant, $\w_n=(2n-1)\pi T$ the Matsubara frequency, and $\phi_\theta=\cos2\theta$ the d-wave form factor.
\footnote{Exactly speaking, the frequency summation should be bounded with a soft-cutoff (with the form of a boson propagator) instead of the hard-cutoff used here. However, only the latter can be mapped to the momentum cutoff strategy as in the BCS treatment.}
In numerical calculations, we use $\lambda=0.3$ and $\Omega=800$, which gives $T_{c0}=1.134\Omega\me^{-2/\lambda}=1.15$ for clean superconductors (with this setup, $0.87T_{c0}$ is taken as the energy unit).
By letting $\Delta_0\to0$, we obtain the generalized $T_c$-formula
\begin{align} \label{eq:Tc}
\ln\frac{T_{c0}}{T_c}=\frac{1}{\pi}\int\md\theta \phi_\theta^2\left[\psi\left(\frac12+\frac{\Gamma_\theta}{2\pi T_c}\right)-\psi\left(\frac12\right)\right]
\end{align}
where $\psi(z)$ is the digamma function.
The results of $T_c$ vs $\Gamma_s$ and $\Gamma_d$ are shown in Fig.~\ref{fig:dos}(e). A stronger $\Gamma_d$ is needed to kill superconductivity as the low energy excitations are less affected. In Fig.~\ref{fig:dos}(f), we show the results of $2\Delta_0/T_c$. It is interesting to find that the anisotropic scattering drives the ratio farther away from the BCS prediction $4.28$ in the clean limit to values as large as $\sim10$. In the literature, a large gap-$T_c$ ratio is often taken as an indication of strong coupling superconductivity. Here, the anisotropic scattering gives another interpretation.

\emph{Superfluid density}.---
After recognizing the clean-dirty dichotomy caused by oxygen vacancies at the single particle level, we turn to discuss the superfluid density, which can be obtained as \cite{Coleman2015}
\begin{align}
\rho_s=e^2\+Nv_F^2T\sum_{\w_n}\int \md\theta \frac{\Delta_\theta^2\cos^2\theta}{\left(\w_n^2+2|\w_n|\Gamma_{\theta}+\Gamma_{\theta}^2+\Delta_\theta^2\right)^{3/2}}
\end{align}
where $v_F$ is the Fermi velocity.
Here, the current vertex correction can be proven to vanish in the non-crossing approximation as a result of the factorized scattering potential (Eq.~\ref{eq:scatteringV}), as shown in the Supplementary Material \cite{sm}.
In Fig.~\ref{fig:superfluid}(a) and (b), $\rho_s$ vs $T$ is plotted for pure $\Gamma_s$ and $\Gamma_d$ scatterings, respectively. The most obvious difference is that any nonzero $\Gamma_s$ causes power law temperature dependence of $\rho_s(T)$, but $\Gamma_d$ preserves the linear dependence as in the clean limit. This is already anticipated from the DOS feature in Fig.~\ref{fig:dos}(b) because the normal fluid density $\rho_n$ is roughly determined by quasiparticles and thus satisfies $\rho_n(T)-\rho_n(0)\propto T$, leaving the superfluid density satisfying $\rho_s(0)-\rho_s(T)=\rho_n(T)-\rho_n(0)\propto T$.

\begin{figure}
\includegraphics[width=1.0\linewidth]{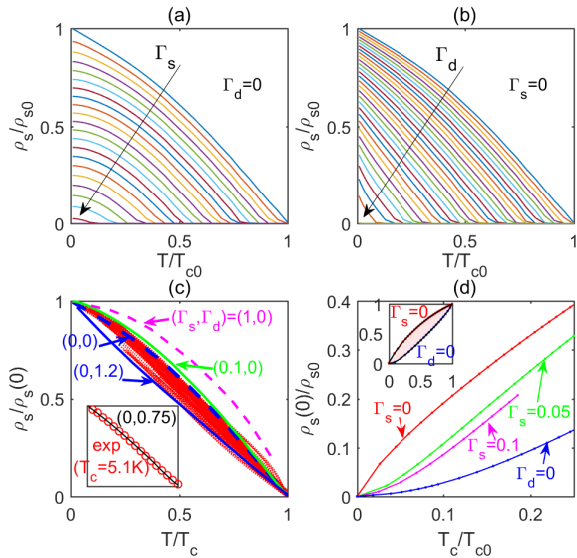}
\caption{(a) Superfluid density $\rho_s/\rho_{s0}$ as a function of $T/T_{c0}$ in the case of isotropic scattering ($\Gamma_d=0$), {for various values of $\Gamma_s$ which increases uniformly from top to bottom.} Here $\rho_{s0}=\rho_s(0)$ in the clean limit. (b) The same as (a) but in the case of anisotropic scattering ($\Gamma_s=0$) and for various values of $\Gamma_d$. (c) $\rho_s/\rho_s(0)$ as a function of $T/T_c$ for $(\Gamma_s,\Gamma_d) = (0,0)$ (dashed blue), $(0.1,0)$ (green), $(1,0)$ (dashed pink), $(0,1.2)$ (blue). The red dots show all $100$ groups of experimental data \cite{Bozovic2016} for comparison. The inset shows the fit to the experimental data with $T_c=5.1$K using $(\Gamma_s,\Gamma_d)=(0,0.75)$. (d) $\rho_s(0)/\rho_{s0}$ as a function of $T_c/T_{c0}$. {The lines are obtained by fixing one of $(\Gamma_s,\Gamma_d)$ (as indicated) while varying the other.} The inset shows two similar curves by fixing $\Gamma_s=0$ and $\Gamma_d=0$, respectively, in a broader scale. The results of all other cases (not shown) fall within the shaded regime bounded by these two curves. {In (c) and (d), the insets share the same axis labels as for the main panels.}}
\label{fig:superfluid}
\end{figure}

In real materials, both $\Gamma_s$ and $\Gamma_d$ are expected to coexist. In order to make quantitative comparisons with the experiment, we renormalize $T$ and $\rho_s$ by $T_c$ and $\rho_s(0)$, respectively, as shown in Fig.~\ref{fig:superfluid}(c). We have superposed the experimental data of all $100$ samples with $T_c$ ranging from $41.6$ to $5.1$K \cite{Bozovic2016}, shown as the red dots. Four typical theoretical curves are plotted: $(\Gamma_s,\Gamma_d)=(0,0)$, $(0.1,0)$, $(1,0)$, and $(0,1.2)$. As can be seen, almost all the experimental data reside within the region enclosed by the curves of $(0.1,0)$ (green line) and $(0,1.2)$ (blue line), while if $\Gamma_s$ is large and dominant, as the $(1,0)$-line shows, the renormalized $\rho_s$-$T$ curve bends more significantly, which is at odds with the experimental data. These observations indicate the more important role of $\Gamma_d$. As an example, the experimental data (open circles) with $T_c=5.1$K are shown in the inset, which is clearly very linear. We can fit the data with pure $\Gamma_d=0.75$ (line) quite well.

Another key observation of the experiment \cite{Bozovic2016} is the linear relationship between $\rho_s(0)$ vs $T_c$. To see that, we plot our results in Fig.~\ref{fig:superfluid}(d). Pure $\Gamma_s$ and $\Gamma_d$ scatterings correspond to blue and red curves, which enclose the physical region (shaded region in the inset), in which $\rho_s(0)$ is always roughly proportional to $T_c$ although not exactly. Therefore, this is a hallmark of the dirty BCS superconductors regardless of isotropic ($\Gamma_s$) or anisotropic ($\Gamma_d$) scatterings. Furthermore, we have also fix $\Gamma_s=0.05$ and $0.1$ and tune $\Gamma_d$ to suppress $T_c$. Both of them show the bending behavior near $T_c\to0$, showing much better behavior than the separate scattering cases in view of the experimental data .

\begin{figure}
\includegraphics[width=1.0\linewidth]{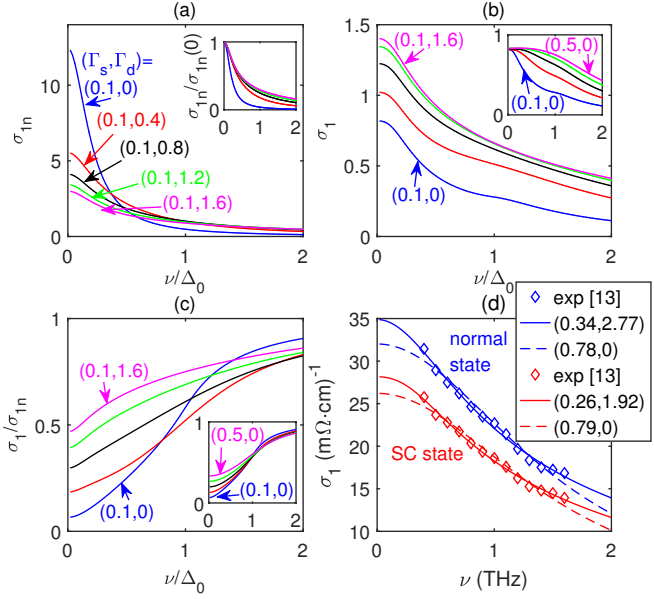}
\caption{(a) Optical conductivity $\sigma_{1n}$ versus frequency $\nu$ in the normal state for various cases of $(\Gamma_s,\Gamma_d)$ (as indicated) in unit of $\Delta_0$. The inset is a replot of $\sigma_1$ in the main panel but normalized by $\sigma_{1n}(0)$. (b) The same as (a) but for the superconducting state conductivity $\sigma_{1}$. The inset shows the result for isotropic scattering, revealing the universal conductivity in the zero frequency limit. (c) $\sigma_1/\sigma_{1n}$ versus $\nu$, with $\sigma_{1n}$ and $\sigma_1$ from (a) and (b), respectively. The inset is for isotropic scattering. In (a)-(c), the line color is associated with the scattering parameter setting identically, and in (b) and (c), the insets share the same axis labels as for the main panels. (d) Best fits (lines) to the experimental data (symbols) with $T_c=17.5$K measured at $22$K (blue) and $1.6$K (red) \cite{Mahmood2019}. The dashed lines are best fits with $\Gamma_s$ alone, and the solid lines include both $\Gamma_s$ and $\Gamma_d$.}
\label{fig:optical}
\end{figure}

\emph{Optical conductivity}.---
From the Kubo formula, the optical conductivity $\sigma_1(\nu)$ can be obtained as \cite{Hirschfeld1989}
\begin{align}\label{eq:optical}
\sigma_1(\nu)=&\frac{\+Ne^2v_F^2}{2\pi^3}\int\md\varepsilon\md\theta\md\w\frac{f(\w)-f(\w+\nu)}{\nu}\cos^2\theta \nn\\
&\text{tr}\left[\text{Im}G(\w,\varepsilon,\theta)\text{Im}G(\w+\nu,\varepsilon,\theta)\right]
\end{align}
where $f(\w)$ is the Fermi distribution function, $G(\w,\varepsilon,\theta)$ is the Green's function in the Nambu space: $G^{-1}=(\w+i\Gamma_\theta)\sigma_0-\varepsilon\sigma_3-\Delta_0\phi_\theta\sigma_1$ with $\sigma_{0,1,3}$ the identity and Pauli matrices. By numerical calculations, we obtain the frequency dependence of $\sigma_1(\nu)$ in both normal and superconducting states (with $\Delta_0=1$ as the energy unit), as shown in Fig.~\ref{fig:optical}(a) and (b), respectively. In the normal state, as $\Gamma_d$ increases, $\sigma_1(\nu)$ becomes more and more broadened to show long tail behavior. This is an important feature obtained in the ``cold spot'' model \cite{Ioffe1998}. In the superconducting state, a nonzero $\Gamma_d$ breaks the universal conductivity \cite{Lee1993}, which applies only in the case of pure $\Gamma_s$ scattering as shown in the inset of Fig.~\ref{fig:optical}(b). Furthermore, we renormalize the superconducting state $\sigma_1(\nu)$ by the normal state value $\sigma_{1n}(\nu)$ to obtain the results shown in Fig.~\ref{fig:optical}(c). The resulted curves show similar behavior to the DOS at low frequencies: nearly linear $\nu$-dependence if $\Gamma_d$ dominates which is quite different from the case of pure $\Gamma_s$ scattering (inset) with quadratic $\nu$-dependence. This can be used to examine the types of scattering in future experiments.

The most obvious feature when $\Gamma_d$ dominates is the behavior of $\sigma_1(\nu)$ is no longer a simple Lorentian-like Drude peak (as in the case of isotropic scattering), since the quasiparticles experience angle dependent scattering rates. In fact, the conductivity should be an integration over a continuous distribution of Lorentzian components. Since the Lorentzian is sharper and higher for smaller scattering rates, the overall line shape of $\sigma_1(\nu)$ should be steeper as the frequency approaches zero. Therefore, if the finite frequency data are used to perform the fit to a single Lorentzian-like Drude peak, as in the THz experiment, the extrapolated value of $\sigma_1(\nu\to0)$ will be lower than the actual one, leading to superficial ``missing'' of the Drude weight.
As an example, in Fig.~\ref{fig:optical}(d) we fit the experimental data with  $T_c=17.5$K (symbols) in both normal (blue) and superconducting (red) states \cite{Mahmood2019}.
The best fits with only $\Gamma_s$ (dashed lines) are found to underestimate $\sigma_1(\nu\to0)$ and the spectral weight, as compared to the result with both $\Gamma_s$ and $\Gamma_d$ (solid lines). Since the pairing gap $\Delta_0$ is not available in the experiment, we take it also as a parameter, and the fitted value is $1.11$THz.
More systematic fitting for a series of overdoped samples can be found in the Supplemental Material \cite{sm}, which supports our main point that apical oxygen vacancy increases with overdoping, leading to larger $\Gamma_d$ scattering.

\emph{Summary and discussions}.---
In summary, we have found that the apical oxygen vacancies give rise to an anisotropic scattering rate $\Gamma_d\cos^2(2\theta)$ which causes a clean-dirty dichotomy for low-high energy excitations. This provides a natural explanation of the anomalous behavior of the superfluid density and THz optical conductivity in the experiments. Therefore, we conclude that the superconducting states of overdoped cuprates are still captured by the BCS theory, as long as the anisotropic scattering rate is adequately considered.

Finally, we make some remarks.
{
(1) In this work, we have omitted many material-dependent details (such as specific tight-binding models, pairing interactions and interaction effects) in order to obtain universal results. The doping dependence enters the problem via the scattering rate $\Gamma_d$, which increases with doping.}
(2) The anisotropic scattering rate has been reported in angle-dependent magnetoresistivity experiments in Nd-LSCO \cite{ADMR2020} and Tl$_2$Ba$_2$CuO$_{6+\delta}$ \cite{ADMR2006}. But to the best of our knowledge, there has not been such report in overdoped LSCO films. A systematic study of the anisotropic scattering rate can check our model and finally answer the question on whether overdoped cuprates are dirty BCS superconductors or not.
(3) Our prediction suggests that it is necessary to extend the optical measurements down to even lower frequency (e.g. GHz) in order to obtain accurate behavior of the optical conductivity.
(4) About superfluid density, there are two other kinds of scalings reported in literature: Uemura's law \cite{Uemura1989} $\rho_s(0)\propto T_c$ and Homes' law $\rho_s(0)\propto\sigma_{dc}T_c$ \cite{Homes2004}. The former is obtained in underdoped cuprates, where phase fluctuations are very strong and cannot be neglected. \cite{Emery1995}. On the other hand, Homes' law can be explained by dirty BCS superconductors with pair-preserving impurities only \cite{AGD1963,Kogan2013}. Here, in the overdoped LSCO films, the oxygen vacancies are pair-breaking, hence Homes' law does not apply.
(5) Recently, the apical oxygen vacancies have also been observed in optimally doped YBa$_2$Cu$_3$O$_{7-x}$ \cite{YBCO2019}. Then, according to our study, $\sigma_1(\nu)$ at low frequency should be non-standard Drude-like, which may be consistent with the early microwave experiment \cite{Turner2003}.
Furthermore, if the apical oxygen vacancies are widely present in different families of cuprates, then the effect of $\Gamma_d$ scattering should be considered carefully in future studies.

\vspace{10pt}
D. W. thanks Congjun Wu, Qi Zhang, Tao Li, Yuan Wan for helpful discussions during the past few years. D. W. also acknowledges Ivan Bozovic for his encouragement on studying their experiments.
This work is supported by National Natural Science Foundation of China (under Grant Nos. 11874205, 11574134, and 12074181), National Key Projects for Research and Development of China (Grant No.2021YFA1400400), Fundamental Research Funds for the Central Universities (Grant No. 020414380185), and Natural Science Foundation of Jiangsu Province (No. BK20200007).

\bibliography{Ovacancy}

\begin{widetext}

\begin{center}
{\LARGE Supplementary Materials}
\end{center}

In this supplementary material, we first derive the scattering potential caused by the apical oxygen vacancies and then the anisotropic scattering rate $\Gamma_d\cos^2(2\theta)$ within the Born approximation. The effects of this scattering rate on some superconducting properties, including density of states, gap, $T_c$, superfluid density, and optical conductivity, are derived in a self-contained way. Finally, some further discussions are given, including first principle calculations to determine the position of the oxygen vacancies (apical or planar), the effect of planar oxygen vacancies (if they exist), and the current vertex corrections.


\section{Disorder model of apical oxygen vacancies}

\subsection{Scattering potential}
At first, suppose there is no oxygen vacancy, we have the following translational invariant Hamiltonian
\begin{align}
H_{dp}=&-\sum_{\av{ij}}(t_{ij}d_{i}^\dag d_j+h.c.)-\sum_{i}t_o( p_i^\dag d_{i+\hat{x}}+p_i^\dag d_{i-\hat{x}}-p_i^\dag d_{i+\hat{y}}-p_i^\dag d_{i-\hat{y}}+h.c.) \nn\\
&+\sum_i(\varepsilon_d d_i^\dag d_i+\varepsilon_o p_i^\dag p_i)
\end{align}
where $d_i^\dag$ generates the in-plane electrons effectively at the Cu-site and $p_i^\dag$ generates the ones at the apical O-site.
The sign change of the out-of-plane hopping (with amplitude $t_o$) between $\pm\hat{x}$ and $\pm\hat{y}$ are caused by the $(x^2-y^2)$ orbital character of the in-plane carriers (either $d_{x^2-y^2}$ orbital electrons or Zhang-Rice singlets), see Fig. 1 in the main text for clarity. We have omitted the spin indices here for simplicity because they are irrelevant to obtain the scattering potential. After integrating out the $p$-electrons (or by a second order perturbation), we obtain
\begin{align}
H_d=-\sum_{\av{ij}}(t_{ij}d_{i}^\dag d_j+h.c.)-\kappa\sum_i\sum_{\delta=\pm\hat{x},\pm\hat{y}} f_\delta f_{\delta'} d_{i+\delta}^\dag d_{i+\delta'}+\sum_i\varepsilon_d d_i^\dag d_i
\end{align}
where $\kappa=\frac{t_o^2}{\varepsilon_o}$, $f_\delta=1$ for $\delta=\pm\hat{x}$ and $f_\delta=-1$ for $\delta=\pm\hat{y}$. The Fourier transformation then gives the band dispersion
\begin{align}
\varepsilon_k=\varepsilon_{kd}-4\kappa(\cos k_x-\cos k_y)^2
\end{align}
where the first term $\varepsilon_{kd}$ is from the pure $d$-electrons, and the second term comes from the coupling to apical oxygen orbitals. In together, $\varepsilon_k$ is taken as the free band without any disorder.

Next, let us consider one apical oxygen vacancy at the origin. Relative to the translational invariant background, the single vacancy is expressed by the impurity Hamiltonian
\begin{align}
H_{\rm imp}=\kappa\sum_{\delta=\pm\hat{x},\pm\hat{y}} f_\delta f_{\delta'} d_{\delta}^\dag d_{\delta'}
\end{align}
whose Fourier transformation then gives a scattering potential
\begin{align}
H_{\rm imp}=\frac{4\kappa}{N}\sum_{kk'}(\cos k_x-\cos k_y)(\cos k_x'-\cos k_y')d_{k}^\dag d_{k'}
\end{align}
which is Eq.~1 in the main text.

\subsection{Anisotropic scattering rate: Born approximation}
As explained in the main text, since the impurity is out-of-plane, its intensity is expected to be quite small such that the Born approximation is acceptable.

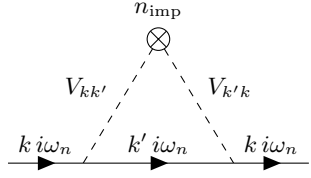
\begin{figure}[h]
\begin{tikzpicture}
    \begin{feynman}
    \vertex (i) ;
    \vertex[right=1cm of i] (a);
    \vertex[right=2cm of a] (b);
    \vertex[right=1cm of b] (f);
    \vertex[right=1cm of a] (c);
    \node[above=1.5cm of c, crossed dot, label=90:\(n_{\rm imp}\)] (d);
    \diagram*{
        (i) -- [fermion, edge label=$k\,i\w_n$] (a) -- [fermion, edge label=$k'\,i\w_n$] (b) -- [fermion, edge label=$k\,i\w_n$] (f),
        (a) -- [scalar, edge label=$V_{kk'}$] (d) -- [scalar, edge label=$V_{k'k}$] (b),
    };
    \end{feynman}
    \end{tikzpicture}
\caption{The single particle self energy in the Born approximation. \label{fig:Born}}
\end{figure}

The self energy under the Born approximation \cite{AGD1963} as shown in Fig.~\ref{fig:Born} is
\begin{align}
\Sigma(k,i\w_n)=n_{\rm imp} \int \frac{\md^2k}{(2\pi)^2} |V_{kk'}|^2 G(k',i\w_n)
\end{align}
where $G(k,i\w_n)=(i\w_n-\varepsilon_k)^{-1}$. As usual, for simplicity, we adopt the wide band approximation with circular Fermi surface and keep the form factor $(\cos k_x-\cos k_y)$ with only the angle dependence $2\cos(2\theta)$, we obtain
\begin{align}
\Sigma(\theta,i\w_n)&=256\kappa^2n_{\rm imp}\+N\cos^2(2\theta) \int \md\varepsilon \int \frac{\md\theta'}{2\pi} \frac{\cos^2(2\theta')}{i\w_n-\varepsilon} \nn\\
&=-i128\pi\kappa^2n_{\rm imp}\+N\text{sgn}(\w_n)\cos^2(2\theta)
\end{align}
by analytic continuation, it gives the desired scattering rate $\Gamma_\theta=\Gamma_d\cos^2(2\theta)$ with $\Gamma_d=128\pi\kappa^2n_{\rm imp}\+N$. Together with the usual isotropic scattering rate $\Gamma_s$, we get Eq.~2 in the main text.

\section{Disorder effect}

\subsection{Density of states}
The retarded Green's function in the superconducting state can be written in the Nambu space as $G_R^{-1}(\w,k)=(\w+i\Gamma_k)\sigma_0-\varepsilon_k\sigma_3-\Delta_k\sigma_1$, where $\sigma_{0,1,3}$ are identity and Pauli matrices. Keeping only the angle dependence and after integration over $\varepsilon$, we obtain
\begin{align}
G_R(\w,\theta)&=\+N\int\md\varepsilon G_R(\w,\theta,\varepsilon) \nn\\
&=\frac{\+N}{2}\int\md\varepsilon\left(\frac{1}{\w+i\Gamma_\theta-E_\theta}+\frac{1}{\w+i\Gamma_\theta+E_\theta}\right)\sigma_0
\end{align}
The angle-dependent partial density of states (PDOS) $\rho(\w,\theta)$ is defined as the imaginary part of $G_R(\w,\theta)$, hence, given by
\begin{align}
\rho(\w,\theta)=-\frac{2}{\pi}\text{Im}[G_{R}(\w,\theta)]_{11}=\frac{\+N}{\pi}\int\md\varepsilon \left[ \frac{\Gamma_\theta}{(\w-E_\theta)^2+\Gamma_\theta^2} + \frac{\Gamma_\theta}{(\w+E_\theta)^2+\Gamma_\theta^2}\right]
\end{align}
which is Eq.~3 in the main text. The angle integral of $\rho(\w,\theta)$ then gives the total density of states (DOS) $\rho(\w)=\int\frac{\md\theta}{2\pi}\rho(\w,\theta)$.

It is interesting to notice that near the nodal region $\theta=\pi/4-\delta\theta$, the d-wave pairing $\Delta_\theta=\Delta_0\cos(2\theta)\approx\Delta_0\delta\theta$ vanishes linearly with $\delta\theta$, while the $\Gamma_d$-scattering vanishes quadratically $\Gamma_\theta\approx\Gamma_d\delta\theta^2$. Therefore, the $\Gamma_d$-scattering has little effect on the low energy quasiparticles, which is quite different from the isotropic scattering and causes the clean-dirty dichotomy as we discussed in the main text.

\subsection{Gap and $T_c$}
Within the BCS framework, we assume a pairing interaction
\begin{align}
H_I=-V\sum_{kk'}\phi_k \phi_{k'} d_{k\up}^\dag d_{-k\dn}^\dag d_{-k'\dn}d_{k'\up}
\end{align}
where $\phi_k=(\cos k_x-\cos k_y)/2$, correspondingly $\phi_\theta=\cos(2\theta)$. Then, the gap function $\Delta_k=\Delta_0\phi_k$ is determined by the self-consistent condition, \begin{align}
\Delta_{k'}&=-\frac{VT\phi_{k'}}{N}\sum_{\w_n,k}\phi_kG_{21}(\w_n,k) \nn\\
&=\frac{2VT\phi_{k'}}{N}\sum_{\w_n>0,k}\frac{\Delta_0 \phi_k^2}{(\w_n+\Gamma_k)^2+\varepsilon_k^2+\Delta_0^2\phi_k^2} \nn\\
&=2\pi \phi_{k'}V\+NT\sum_{\w_n>0}\int\frac{\md\theta}{2\pi}\frac{\Delta_0\phi_\theta^2}{\sqrt{(\w_n+\Gamma_\theta)^2+\Delta_0^2\phi_\theta^2}}
\end{align}
which gives Eq.~4 in the main text by defining $\lambda=V\+N$. In the last line, we have also adopted the wide band approximation with circular Fermi surface.

Next, by letting $\Delta_0\to0$, we can determine $T_c$. The calculation is in parallel to Ref.~\onlinecite{AG1961}.
The above self-consistent equation becomes
\begin{align}
1&=2\pi\lambda T_c\sum_{\w_n>0}^{\Omega}\int\frac{\md\theta}{2\pi}\frac{\phi_\theta^2}{\w_n+\Gamma_\theta} \nn\\
&=2\pi\lambda T_c\sum_{\w_n>0}^{\Omega}\int\frac{\md\theta}{2\pi}\left( \frac{\phi_\theta^2}{\w_n+\Gamma_\theta}-\frac{\phi_\theta^2}{\w_n}+\frac{\phi_\theta^2}{\w_n} \right) \nn\\
&=\lambda\sum_{n=0}^{\infty}\int\frac{\md\theta}{2\pi}\phi_\theta^2\left( \frac{1}{n+\frac12+\frac{\Gamma_\theta}{2\pi T_c}}- \frac{1}{n+\frac12}\right)+2\pi \lambda T_c\sum_{\w_n>0}^\Omega \int\frac{\md\theta}{2\pi}\frac{\phi_\theta^2}{\w_n} \label{eq:gap-eq}
\end{align}
where we have added and subjected the same term (with $\Gamma_\theta=0$) such that the summation of the first two terms converges and upper limit can now be pushed to infinity. The last term is the same as the clean case without disorder except $T_{c0}$ replaced by $T_{c}$. Using the clean case result $T_{c0}=\alpha\Omega\me^{-2/\lambda}$ for d-wave pairing, we obtain
\begin{align}
\frac{1}{\lambda}-2\pi T_c\sum_{\w_n>0}^\Omega\int\frac{\md\theta}{2\pi}\frac{\phi_\theta^2}{\w_n}=\frac12\ln\frac{T_c}{T_{c0}}
\end{align}
substituting it into Eq.~\ref{eq:gap-eq}, we obtain
\begin{align}
\ln\frac{T_{c0}}{T_c}=\int\frac{\md\theta}{\pi}\phi_\theta^2\left[\psi\left(\frac12+\frac{\Gamma_\theta}{2\pi T_c}\right) - \psi\left(\frac12\right)\right]
\end{align}
which is Eq.~5 in the main text.

\subsection{Superfluid density}
Next, let us consider the superfluid density. The calculation mainly follows Ref.~\onlinecite{Coleman2015}. The superfluid density is given by two diagrams as shown in Fig.~\ref{fig:superfluid}. The first diagram is from the paramagnetic current $J_k^P=e\nabla\varepsilon_k d_k^\dag d_k$ and the second one is from the diamagnetic current $J_k^d=e^2\nabla^2 \varepsilon_k d_k^\dag d_k$.

\begin{figure}[h]
\begin{tikzpicture}
    \begin{feynman}
    \vertex (i) ;
    \node[right=1cm of i,dot, label=0:$v_k\sigma_0$] (a);
    \node[right=2.5cm of a,dot,label=180:$v_k\sigma_0$] (b);
    \vertex[right=1cm of b] (f);
    \diagram*{
        (i) -- [photon] (a),
        (a) -- [fermion,half left,edge label=$k\,i\w_n$] (b),
        (b) -- [fermion,half left,edge label=$k\,i\w_n$] (a),
        (f) -- [photon] (b)
        };
    \end{feynman}
    \end{tikzpicture}
\quad\quad\quad
\feynmandiagram [layered layout] {
  a -- [photon] b [dot] -- [fermion,out=135, in=45, loop, min distance=4cm, edge label=$k\,i\w_n$] b [label=270:$\nabla^2\varepsilon_k\sigma_3$] -- [photon] c,
};
\caption{Feynman diagrams for the superfluid density $\rho_s$. \label{fig:superfluid}}
\end{figure}
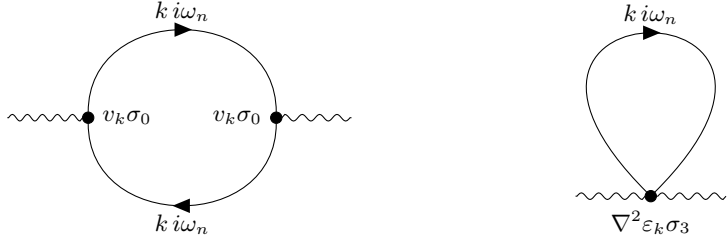

Put these two contributions together, we obtain the superfluid density
\begin{align}
\rho_s&=e^2\frac{T}{N}\sum_{\w_n,k}\text{Tr}\left[G(k,i\w_n)v_{k,x}G(k,i\w_n)v_{k,x}+G(k,i\w_n)(\nabla_xv_k)\tau_3\right] \nn\\
&=4e^2\frac{T}{N}\sum_{\w_n,k} \frac{v_{k,x}^2 \Delta_k^2}{[(|\w_n|+\Gamma_k)^2+\varepsilon_k^2+\Delta_k^2]^2}
\end{align}
where we keep only the $xx$-component. Similar to previous sections, the $k$-summation is replaced by integrals over $\varepsilon$ and $\theta$ under wide band approximation with circular Fermi surface, giving rise to
\begin{align}
\rho_s&=4e^2T\+N v_F^2\sum_{\w_n} \int\frac{\md\theta}{2\pi} \int \md\varepsilon\frac{\phi_\theta^2\Delta_\theta^2}{[(|\w_n|+\Gamma_\theta)^2+\varepsilon^2+\Delta_\theta^2]^2} \nn\\
&=e^2T\+Nv_F^2\sum_{\w_n}\int\md\theta \frac{\phi_\theta^2\Delta_\theta^2}{[(|\w_n|+\Gamma_\theta)^2+\Delta_\theta^2]^{3/2}}
\end{align}
which is Eq.~6 in the main text.

\subsection{Optical conductivity}
The optical conductivity is contributed only by the paramagnetic current as shown in Fig.~\ref{fig:optical}.
\begin{figure}[h]
\begin{tikzpicture}
    \begin{feynman}
    \vertex (i) ;
    \node[right=1cm of i,dot,label=0:$v_{k}\sigma_0$] (a);
    \node[right=2.5cm of a,dot,label=180:$v_{k}\sigma_0$] (b);
    \vertex[right=1cm of b] (f);
    \diagram*{
        (i) -- [photon,edge label=$i\nu_n$] (a),
        (a) -- [fermion,half left,edge label=$k\,i\w_n+i\nu_n$] (b),
        (b) -- [fermion,half left,edge label=$k\,i\w_n$] (a),
        (b) -- [photon,edge label=$i\nu_n$] (f)
        };
    \end{feynman}
    \end{tikzpicture}
\caption{Feynman diagrams for the optical conductivity. \label{fig:optical}}
\end{figure}
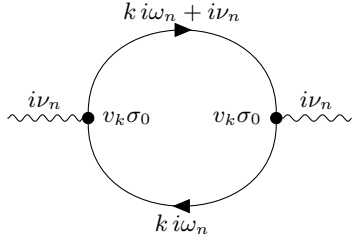
The current-current correlation function $K(i\nu_n)$ is given by
\begin{align}
K(i\nu_n)&=-e^2\frac{T}{N}\sum_{\w_n,k}\text{Tr}\left[G(k,i\w_n)v_{k,x}G(k,i\w_n+i\nu_n)v_{k_x}\right] \nn\\
&=-e^2\frac{T}{N}\sum_{\w_n,k}v_{k,x}^2\int\md\w \int\md\w'\frac{\text{Tr}\left[\text{Im}G(k,\w)\text{Im}G(k,\w')\right]}{(i\w_n-\w)(i\w_n+i\nu_n-\w')}
\end{align}
where the spectral representation has been used. Now the summation of $i\w_n$ can be completed and gives
\begin{align}
K(i\nu_n)=-\frac{e^2}{\pi^2N}\sum_{k}v_{k,x}^2\int\md\w\int\md\w'\frac{f(\w)-f(\w')}{i\nu_n+\w-\w'}\text{Tr}\left[\text{Im}G(k,\w)\text{Im}G(k,\w')\right]
\end{align}
After analytic continuation $i\nu_n\to\nu+i0^+$, the optical conductivity $\sigma_1(\nu)$ can be obtained as
\begin{align}
\sigma_1(\nu)&=\text{Re}\left[\frac{K(i\nu_n\to\nu+i0^+)}{-i\nu}\right] \nn\\
&=\frac{e^2}{\pi^2N}\sum_k v_{k,x}^2 \int\md\w \frac{f(\w)-f(\w+\nu)}{\nu}\text{Tr}\left[\text{Im}G(k,\w)\text{Im}G(k,\w+\nu)\right]
\end{align}
Finally, we replace the $k$-summation by energy and angle integrals and arrive at the final expression
\begin{align} \label{eq:optical}
\sigma_1(\nu)=\frac{e^2v_F^2\+N}{2\pi^3}\int\md\varepsilon\int\md\w\int\md\theta \cos^2\theta \frac{f(\w)-f(\w+\nu)}{\nu}\text{Tr}\left[\text{Im}G(\varepsilon,\theta,\w)\text{Im}G(\varepsilon,\theta,\w+\nu)\right]
\end{align}
which is Eq.~7 in the main text.

\begin{figure}[h!]
\centering
\includegraphics[width=0.7\linewidth]{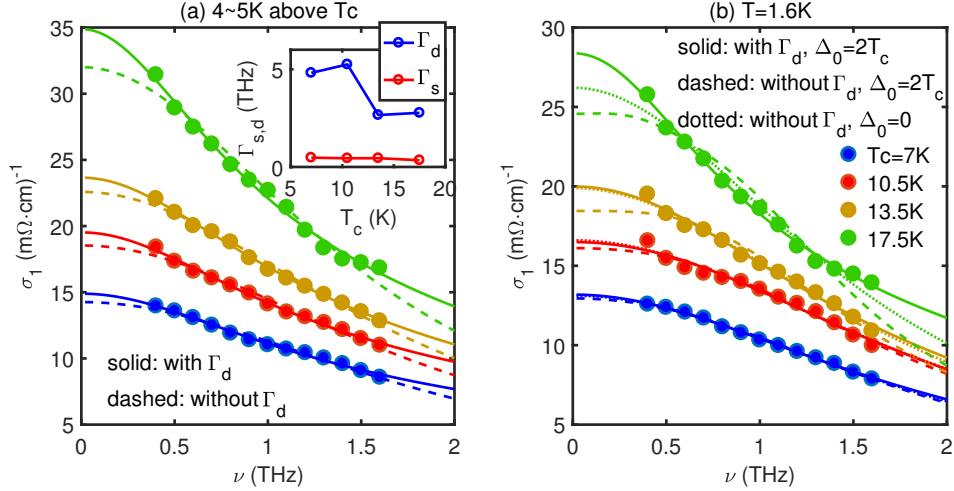}
\caption{
Fittings of the THz experimental data of $\sigma_1$ on four samples \cite{Mahmood2019} in the normal state (a) and superconducting state (b), respectively, with (solid lines) and without (dashed lines) the anisotropic scattering rate $\Gamma_d$. In the superconducting state, we fix $\Delta_0=2T_c$ as a rough estimate of the pairing gap. For comparison, we also present the fitted results (dotted lines) with $\Delta_0=0$ as performed in Ref.~\onlinecite{Mahmood2019}. In the inset of (a), we plot the best fitted parameters $(\Gamma_s,\Gamma_d)$ versus $T_c$, respectively.
}
\label{fig:fitExp}
\end{figure}

On the basis of Eq.~\ref{eq:optical}, we fit the experimental data of $\sigma_1$ from Ref.~\onlinecite{Mahmood2019}. We read the data points from $\nu=0.4$ to $1.6$THz by every $0.1$THz. The best fits are shown in Fig.~\ref{fig:fitExp}(a) for normal state and in Fig.~\ref{fig:fitExp}(b) for superconducting state, respectively, with (solid lines) and without (dashed or dotted lines) $\Gamma_d$.
For the superconducting state, we have fixed $\Delta_0=2T_c$ as a rough estimate of the pairing gap. Compared with pure $\Gamma_s$-fitting, the results using both Gs and Gd agree better to most of the data. As anticipated, the pure-$\Gamma_s$ fitting underestimates both $\sigma_1(\nu\to0)$ and the spectral weight (enclosed area). We further show the best fitted parameters $(\Gamma_s,\Gamma_d)$ obtained from the normal state data in the inset of Fig.~\ref{fig:fitExp}(a), which shows clearly that $\Gamma_d$ becomes stronger upon overdoping (reducing $T_c$) while $\Gamma_s$ barely changes, supporting our main point that the apical oxygen vacancies become more and more important in overdoped LSCO upon doping.

\section{Further discussions}

\subsection{First principle calculations}

In order to determine the position of the oxygen vacancies theoretically, we performed first principle calculations on La$_{2-x}$Sr$_x$CuO$_{4-y}$ within the local density approximation (LDA), which has been found to work very well in the overdoped cuprates \cite{Kramer2019}. By constructing two kinds of super unit cells, $2\times2\times1$ and $2\times2\times2$, with one oxygen vacancy, and optimizing the positions of the doped Sr atoms, we obtain the total energy difference between two kinds of oxygen vacancies $\Delta E=E_{\rm apical}-E_{\rm planar}$ as shown in Fig.~\ref{fig:LDA}(a). Without Sr doping, the planar oxygen vacancy $E_{\rm planar}$ is lower than the apical one $E_{\rm apical}$, but upon Sr doping, their difference reduces and finally reverses at high doping levels. Such a tendency becomes stronger when we enlarge the super unit cell from $2\times2\times1$ to $2\times2\times2$, indicating the oxygen vacancies in overdoped LSCO are very likely to be apical rather than planar ones, in agreement with the Raman experiment \cite{Kim2017}.
This behavior can be roughly understood from the density of states (DOS), as shown in Fig.~\ref{fig:LDA}(b). As can be seen, the planner oxygen vacancy reduces the DOS near the Fermi level at $x=0$ and thus saves more energy. Instead, the apical vacancy enhances the DOS near the Fermi level. Upon hole doping caused by Sr, the DOS of the apical case is reduced and thus finally becomes preferred than the planar case.

\begin{figure}
\centering
\includegraphics[width=0.425\linewidth]{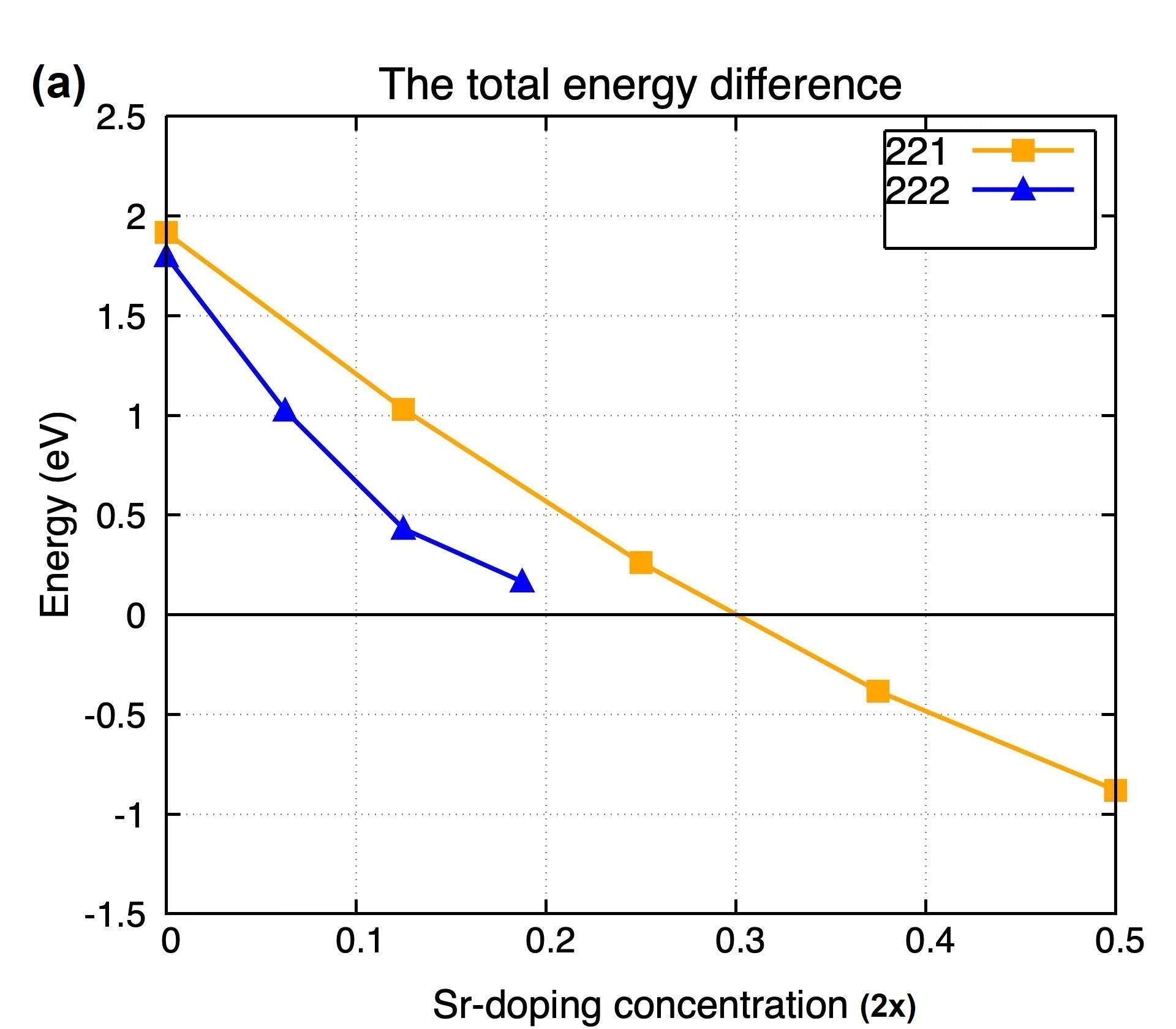}
\includegraphics[width=0.4\linewidth]{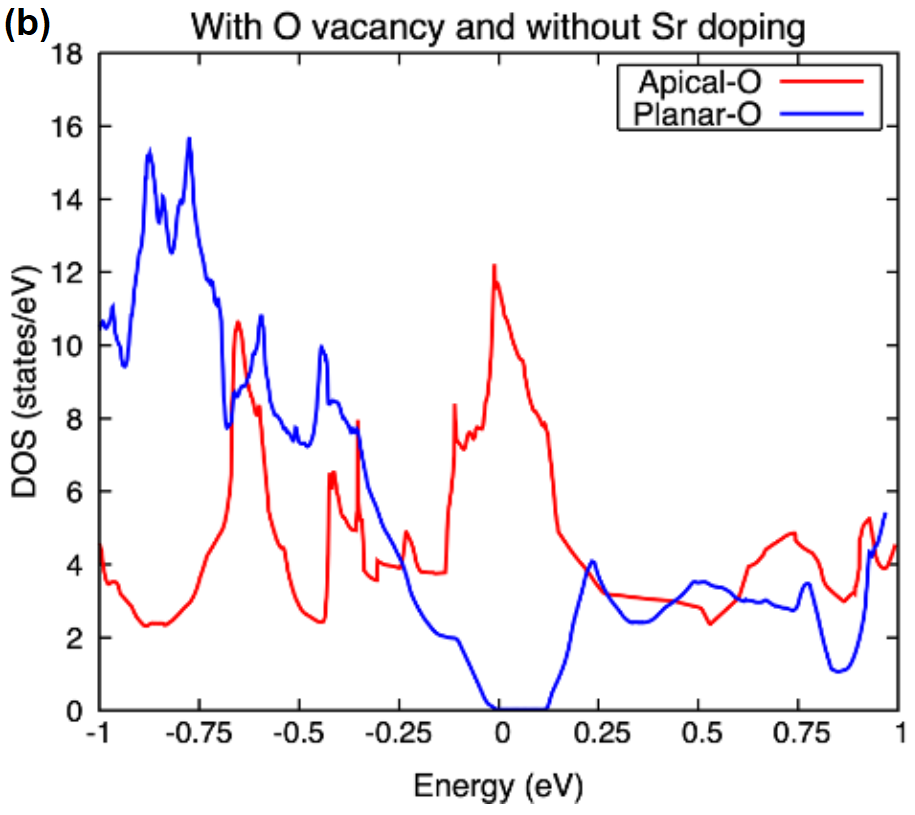}
\caption{LDA results of apical and planar oxygen vacancies. (a) The energy difference $\Delta E=E_{\rm apical}-E_{\rm planar}$ is plotted versus the Sr concentration (2x) with two kinds of super unit cells. (b) The total DOS near the Fermi level at $x=0$ for apical and planar oxygen vacancies, respectively.}
\label{fig:LDA}
\end{figure}

\subsection{Planar oxygen vacancies}
From both the experiment and our LDA calculations, we have found the oxygen vacancies are dominated by apical ones in overdoped LSCO. However, at small doping, the planar ones are more stable. Therefore, we may also ask what happens for the planar oxygen vacancies?

At first, one planar oxygen vacancy breaks the Zhang-Rice singlets on its two neighboring sites. To describe such an effect, we consider a model by adding two onsite potentials given by
\begin{align}
H^{\rm I}_{x,y}=V(d_0^\dag d_0+d_{x,y}^\dag d_{x,y})=\frac{V}{N}\sum_{kk'} \left[1+\me^{i(k_{x,y}'-k_{x,y})}\right]d_k^\dag d_{k'}
\end{align}
where the subscript $x,y$ labels the position of the vacancy is on the $x$-bond or $y$-bond, hence, $d_0^\dag$, $d_{x}^\dag$ and $d_{y}^\dag$ generate electrons on $(0,0)$, $(\hat{x},0)$ and $(\hat{y},0)$, respectively. We call this scattering process as type-I. Its amplitude $|V_{kk'}|^2$ is given by
\begin{align}
|V^{\rm I}_{kk'}|^2=4V^2\cos^2\left(\frac{k_{x,y}'-k_{x,y}}{2}\right) \label{eq:in-plane1}
\end{align}
Such a disorder scattering is difficult to treat exactly as it depends on the transferred momentum $(k_{x,y}-k_{x,y}')$ (but not $k$ and $k'$ independently like the apical case) and breaks the rotational symmetry. Nevertheless, let us qualitatively consider its role on the d-wave superconducting properties. If we keep only the dominant scattering processes, they occur at $k_x\approx k_x'$ (for $x$-bond vacancies)and $k_y\approx k_y'$ (for $y$-bond vacancies). These scatterings preserve the pairing sign and are expected to be not pair-breaking, in some sense like the point disorder in s-wave superconductors. Therefore, the standard nonmagnetic disorder effects on the conventional dirty s-wave superconductors are expected: such disorders should have little effect on $T_c$ but reduce the superfluid density. Furthermore, these scatterings do not vanish for nodal quasiparticles, hence, the scattering rate $\Gamma$ is finite in this region (like $\Gamma_s$) which cannot explain the linear $T$-dependence of the superfluid density.

On the other hand, one planar oxygen vacancy can also change the hopping on this bond (called type-II process), which can be described by the following Hamiltonian
\begin{align}
H^{\rm II}_{x,y}=V'(d_0^\dag d_{x,y}+d_{x,y}^\dag d_{0})=\frac{V'}{N}\sum_{kk'} \left(\me^{ik'_{x,y}}+\me^{-ik_{x,y}}\right)d_k^\dag d_{k'}
\end{align}
The symbols are similar to the above. The amplitude of this scattering potential is
\begin{align}
|V^{\rm II}_{kk'}|^2=4V'^2\cos^2\left(\frac{k_{x,y}'+k_{x,y}}{2}\right) \label{eq:in-plane2}
\end{align}
which is largest at $k_x\approx-k_x'$ (for $x$-bond vacancies) and $k_y\approx-k_y'$ (for $y$-bond vacancies). These scatterings are also not pair-breaking and thus are expected to be similar to the above type-I scatterings.

Put these two processes together, we conclude that the planar oxygen vacancies behave more like pair-preserving disorders in d-wave superconductors and cannot explain the superfluid density experiment.

\subsection{Current vertex correction}
Now let us examine the effect of the current vertex correction caused by the anisotropic disorder scattering. We label the bare current vertex by $v_k$ and the dressed vertex by $\Lambda_k$. They are connected in a self-consistent way as shown in Fig.~\ref{fig:vc}, where we have neglected all crossing diagrams which are at least of order $n_{\rm imp}^2\kappa^4$ and are expected to be more relevant to localization \cite{Lee1985}. In fact, the resistivity in the overdoped LSCO becomes smaller and smaller upon doping, indicating the system are more and more itinerant than localization \cite{Bozovic2016}.
\begin{figure}[h]
\begin{align}
\vcenter{\hbox{\begin{tikzpicture}
  \begin{feynman}
    \vertex (i);
    \vertex [below=3cm of i] (j);
    \vertex [below=1.5cm of i] (k);
    \node [right=1.5cm of k,blob,label=80:$\Lambda_k$] (v);
    \vertex [right=1.5cm of v] (o);
    \diagram*{
      (i) --[fermion, edge label=$k$] (v) -- [fermion, edge label=$k$] (j),
      (v) --[photon] (o),
    };
  \end{feynman}
\end{tikzpicture}}}
~=~
\vcenter{\hbox{\begin{tikzpicture}
  \begin{feynman}
    \vertex (i);
    \vertex [below=3cm of i] (j);
    \vertex [below=1.5cm of i] (k);
    \node [right=1.5cm of k,dot,label=80:$v_k$] (v);
    \vertex [right=1.5cm of v] (o);
    \diagram*{
      (i) --[fermion,edge label=$k$] (v) -- [fermion,edge label=$k$] (j),
      (v) --[photon] (o),
    };
  \end{feynman}
\end{tikzpicture}}}
~+~
\vcenter{\hbox{\begin{tikzpicture}
  \begin{feynman}
    \vertex (i);
    \vertex [below=3cm of i] (j);
    \node [below=1.35cm of i,crossed dot,label=180:$n_{\rm imp}$] (k);
    \node [blob, right=1.5cm of k,label=80:$\Lambda_{k'}$] (v);
    \vertex [right=1.5cm of v] (o);
    \vertex [below=0.5cm of i] (x);
    \vertex [right=0.5cm of x] (i2);
    \vertex [above=0.5cm of j] (y);
    \vertex [right=0.5cm of y] (j2);
    \diagram*{
      (i) --[fermion, edge label=$k$] (i2) -- [fermion, edge label=$k'$] (v) -- [fermion, edge label=$k'$] (j2) --[fermion, edge label=$k$] (j),
      (v) --[photon] (o),
      (j2) -- [scalar,edge label=$V_{kk'}$] (k) -- [scalar, edge label=$V_{k'k}$] (i2),
    };
  \end{feynman}
\end{tikzpicture}}} \nn
\end{align}
\caption{Current vertex correction caused by the disorder scattering. \label{fig:vc}}
\end{figure}
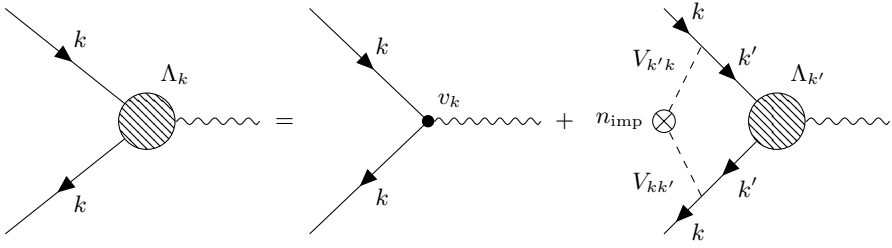

From Fig.~\ref{fig:vc}, we can write down the self-consistent equation
\begin{align}
\Lambda_k=v_k+\frac{n_{\rm imp}}{N}\sum_{k'}|V_{kk'}|^2G^2(i\w_n,k')\Lambda_{k'}
\end{align}
For a general scattering $V_{kk'}$, the above equation can be solved to give rise to an additional factor $(1-\cos\theta_{kk'})$ in the so-called transport scattering rate $\tau_{\rm tr}^{-1}$ which enters $\sigma_1(\nu)$ \cite{AGD1963}. But here, in our model, it becomes quite simple because the disorder scattering potential $V_{kk'}$ is already factorized $V_{kk'}=16\kappa\cos2\theta_k\cos2\theta_{k'}$ such that the summation over $k'$ can be completed exactly. Moreover, notice that $\Lambda_{k'}$ is time reversal ($k'\to-k'$) odd, while $|V_{kk'}|^2G^2(i\w_n,k')$ is time reversal even, the integral over $k'$ must be zero, which means $\Lambda_k=v_k$. Therefore, in our disorder model, the current vertex correction totally vanishes, if the crossing diagrams can be safely neglected.

\end{widetext}

\end{document}